\documentclass[12pt]{article}
\usepackage{authblk}
\usepackage{graphicx}
\usepackage{amssymb}
\usepackage{amsmath}
\title{
Analogue cosmology in a hadronic fluid
 }
\author[1,2]{Neven Bili\'c\thanks{bilic@irb.hr}}
\affil[1]{Rudjer Bo\v skovi\'c Institute, 10002 Zagreb, Croatia}
\author[1]{Dijana Toli\'c\thanks{dijana.tolic@irb.hr}}
\affil[2]{Departamento de F\'isica,
Universidade Federal de Juiz de Fora, 36036-330,
Juiz de Fora, MG, Brazil}
\date{\today}
\begin{document}
\maketitle

The expansion of hadronic matter that takes place immediately after a heavy ion 
collision has certain similarity with the cosmological expansion. We study the analogue geometry 
of the expanding hadronic fluid, using the the formalism of relativistic 
acoustic geometry \cite{bilic3,moncrief,visser2}. 
We show that the propagation of massless pions provides a geometric  analog of expanding spacetime
equivalent to an open  ($k=-1$)  FRW cosmology. 
 Here, we study  general conditions for the formation of a trapped region with
the  inner boundary   as a marginally trapped surface. 

Our approach is based on the linear sigma model combined with a  boost invariant 
Bjorken type spherical expansion.\footnote{The assumption of spherical expansion  is appropriate 
for $e^+e^-$ collisions \cite{cooper} as in this case the jets are produced with no directional preference.}
A Bjorken-type expansion is a simple and very useful hydrodynamic model that 
reflects the boost invariance of the deep inelastic scattering  in high energy 
collisions.\footnote{The original model \cite{bjorken} was introduced to describe the longitudinal expansion only.}

To describe the effective geometry of the expanding hadronic fluid, 
we introduce the analogue gravity metric $G_{\mu\nu}$.
 The dynamics of massless pions is described by the equation of motion for three   
pion fields $\pi^i$ and sigma meson field $\sigma$ propagating in curved space-time. 
The equation of motion is equivalent 
to the d'Alembertian equation of motion for a massless scalar field propagating 
in a (3+1)-dimensional Lorentzian geometry
%
\begin{equation}
\frac{1}{\sqrt{-G}}\,
\partial_{\mu}
(\sqrt{-G}\,
G^{\mu\nu})
\partial_{\nu}
\mbox{\boldmath{$\pi$}}
+V(\sigma,
\mbox{\boldmath{$\pi$}})
\mbox{\boldmath{$\pi$}}=0 ,
\label{eq028}
\end{equation}
where $G_{\mu\nu}$ is the analogue metric tensor and $V(\sigma,${\boldmath$\pi$)}
 is a potential
that describes effective interaction between the mesons.

\subsubsection*{Dynamics of the chiral fluid}
In order to
draw the analogy with cosmology now we consider a spherically symmetric 
Bjorken expansion of the chiral fluid which is invariant under
radial boosts. In this model the radial three-velocity in radial coordinates
$x^\mu=(t,r,\vartheta,\varphi)$ is  a simple function 
$v=r/t$. Then the four-velocity  
 is  given by $u^\mu= (t/\tau, r/\tau,0,0)$, where $\tau=\sqrt{t^2-r^2}$ is the 
proper time. With the substitution $t=\tau \cosh y$, $r=\tau \sinh y$ the radial velocity 
is expressed as $v=\tanh y$ and the four-velocity as $u^\mu=(\cosh y,\sinh y,0, 0)$.
This substitution may be regarded as a coordinate transformation from ordinary radial coordinates to  new coordinates
$(\tau,y,\vartheta,\varphi)$
in which the flat background metric takes the form
\begin{equation}
g_{\mu\nu}= {\rm diag} \left(1,  -\tau^2, -  
\tau^2\sinh^2\! y, - \tau^2\sinh^2\! y \sin^2\!\theta \right),
\label{eq108}
\end{equation}
and the velocity componets become  $u^\mu=(1,0,0,0)$.
Hence,
the new coordinate frame is comoving.
The metric  corresponds to an FRW expanding cosmological model 
with cosmological scale  $a=\tau$ and negative spatial curvature.
 We mapped the spatially flat
Minkowski spacetime into an expanding FRW spacetime with cosmological scale $a=\tau$ and negative spatial 
curvature. The resulting flat spacetime with metric (\ref{eq108}) is  known in cosmology 
as the {\it Milne universe} \cite{milne}.

The temperature of the expanding chiral fluid, to a good approximation, is proportional to $\tau^{-1}$.
This follows from the fact that the chiral matter is dominated by massless pions,
and hence, the density of the fluid 
may be approximated by the density $\rho=(g\pi^2/30) T^4$ of an ideal massless boson gas 
\cite{landau}.
 Using this and the energy-momentum conservation one finds
$T=c_0/\tau$ where the constant $c_0$ may be fixed from 
the phenomenology of high energy collisions. 

\subsubsection*{Dynamics of pions}

The dynamics of pions in the hadronic fluid can be described using a linear sigma model as 
an effective low energy model of strong interactions. The basic model involves four scalar fields (three
pions and a sigma meson)   $\varphi\equiv (\sigma ,$ {\boldmath$\pi$)}  which constitute the 
$(\frac{1}{2},\frac{1}{2})$ representation of the chiral SU(2)$\times$SU(2). 
In the chirally symmetric phase
at temperatures above the chiral transition point the mesons are massive with equal masses.
In the chirally broken phase the pions are massless and 
 sigma meson acquires a nonzero mass proportional to the chiral condensate.

At
temperatures below the chiral phase transition point the pions, although being massless,
propagate slower than light \cite{pisarski2,son1,son2} with 
a velocity approaching zero at the critical
temperature. Hence, it is very likely that there exists a region where the 
flow velocity exceeds the pion velocity and the analogue trapped region may form. 

The dynamics of mesons
 in a medium  is described by
a chirally symmetric Lagrangian of the form
\cite{son1,son2,bilic2}
\begin{equation}
{\cal{L}} =
 \frac{1}{2}(a\, g^{\mu\nu}
 +b\, u^{\mu}u^{\nu})\partial_{\mu} \varphi
 \partial_{\nu} \varphi
 - \frac{m_0^2}{2}
 \varphi^2
- {\lambda\over 4}
 (\varphi^2)^2 ,
 \label{eq1}
\end{equation}
where $u_{\mu}$ is the velocity of the fluid,
and $g_{\mu\nu}$ is the background metric.
The parameters
 $a$ and $b$ depend  on the local temperature $T$
 and on the parameters of the model $m_0$ and $\lambda$
 and may be extracted from the pion self-energy at non-zero temperature. \cite{bilic2}
At zero temperature the medium is  absent in which case $a=1$ and $b=0$. 
Propagation of pions is governed by the equation of motion (\ref{eq028})
with the analogue metric tensor given by
\begin{equation}
G_{\mu\nu} =\frac{a}{c_{\pi}}
[g_{\mu\nu}-(1-c_{\pi}^2)u_{\mu}u_{\nu}],
\label{eq022}
\end{equation}
and the pion velocity squared $c_{\pi}^2=a/(a+b)$. Hence, the pion field propagates in a (3+1)-dimensional
effective geometry described by the metric
$G_{\mu\nu}$. It is convenient to work in comoving coordinates $(\tau,y,\vartheta,\varphi)$
with background metric $g_{\mu\nu}$ defined by (\ref{eq108}).
In these coordinates 
the analogue metric tensor (\ref{eq022}) is diagonal with components
\begin{equation}
G_{\mu\nu}=\frac{a}{c_\pi}{\rm diag} \left(c_\pi^2,  -\tau^2, -  
\tau^2\sinh^2\! y, - \tau^2\sinh^2\! y \sin^2\!\theta \right),
\label{eq008}
\end{equation}
where the parameters $a$ and $c_\pi$ are functions of the temperature $T$ which in turn
is a function of $\tau$. In the following we assume
that these functions are positive.

In contrast to \cite{bilic3}, where it was assumed that both the background geometry and 
 the  flow were stationary, in an expanding fluid the flow 
is essentially time dependent. Hence, 
the acoustic geometry formalism  
 must be adapted to a non-stationary space time.

\subsubsection*{Analogue horizons}

For a relativistic flow in curved spacetime the
apparent and trapping horizons may be defined in the same way as in general relativity. 

The key element in the study of trapped surfaces is the expansion parameter $\varepsilon_\pm$ of  null geodesics.
A two-dimensional surface $S$ with spherical topology is called a {\em trapped surface}  if 
the families of ingoing and outgoing null geodesics normal to the surface are  both converging or both diverging.
More precisely,  the 
 expansion parameters
\begin{equation}
\varepsilon_\pm=\nabla_\mu l_\pm^\mu  
\label{eq244}
\end{equation}
 on a trapped surface $S$ should satisfy
$\varepsilon_+\varepsilon_- >0$.
A two-dimensional surface $H$ is said to be {\em future inner marginally trapped} 
if the future directed null expansions 
 on $H$ satisfy the conditions:  $\varepsilon_+|_H=0$, $l_-^\mu\partial_\mu\varepsilon_+|_H>0$ 
 and $\varepsilon_-|_H<0$.
 We shall refer to this surface  as
the {\em apparent horizon} since it is equivalent to the apparent horizon in cosmological context.

To define the analogue apparent horizon 
we need to examine the behaviour of radial null geodesics of the analogue metric
(\ref{eq008}) in which $a$ and $c_\pi$ are functions of $\tau$. 
Using the geodesic equation $l^\mu \nabla_\mu{l^\nu}=0$, 
from (\ref{eq244}) we find the condition for the apparent horizon
 \begin{equation}
 \frac{1}{v} \pm\frac{\dot{\alpha}}{\beta} =0
\label{eq118}
\end{equation}
where
$\alpha(\tau)=\tau\sqrt{a/c_\pi}$ and $\beta(\tau)=\sqrt{ac_\pi}$. 
This equation defines a hypersurface dubbed the {\em analogue  trapping horizon} and its
solution  determines the location  of 
the analogue apparent horizon $r_H$ as a function of time. From (\ref{eq118}) 
it follows that the region of spacetime $\tanh y \geq |\beta/\dot{\alpha}|$ is trapped. 
Specifically, it is future trapped 
if $\dot{\alpha} < 0$  and past trapped if $\dot{\alpha} > 0$.

Spacetime diagrams corresponding to the metric (\ref{eq008})
are presented in Fig. 1 showing  future directed radial null geodesics.
The origin in the plots in both panels  corresponds to the critical value $\tau_{\rm c}$
at which $c_\pi$ vanishes. At 
$\tau=\tau_{\rm max}$ we have $|\beta/\dot{\alpha}|=1$
so the trapping horizon ends at the point
$\tau=\tau_{\rm max}$, $y=\infty$.
\begin{figure}[t]
\begin{center}
\includegraphics[width=0.5\textwidth,trim= 0 0cm 0cm 0cm]{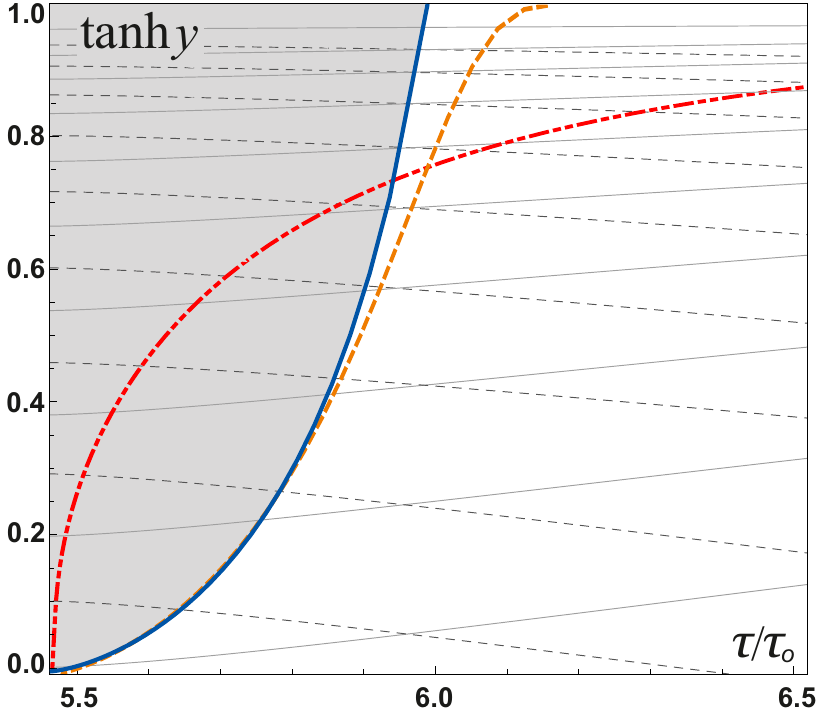}
\caption{Spacetime diagram of outgoing (full line) and ingoing (dashed line)
radial null geodesics in $(\tau,y)$ coordinates. The shaded area represents the evolution of the trapped region.
The trapping horizon is represented by the full bold line
with the endpoint at  $\tau=\tau_{\rm max}=6.0182\tau_0$.
The dashed and dash-dotted bold lines represent the evolution of the analogue and naive Hubble horizons,
respectively.}
 \label{fig3}
\end{center}
\end{figure}

We next examine  the analogue Hubble rate, in particular its
behavior in the neighborhood of the critical point.
For the  spacetime defined by the metric (\ref{eq008}) the Hubble rate
is given by ${\cal H} =\partial_\tau \left(\tau\sqrt{a/c_\pi}\right)/(a\tau)$. We find that ${\cal H}$ is negative for $\tau$ in 
the entire range $\tau_{\rm c} \leq \tau \leq \tau_{\rm max}$ and
scales as ${\cal H} \propto-(\tau-\tau_{\rm c})^{-1.17}$
as $\tau$ approaches $\tau_{\rm c}$.
Hence, our 
 cosmological model  describes
a shrinking FRW universe with a singularity at the critical point.

\subsubsection*{Surface gravity and analogue  Hawking effect}
Next we study the Hawking effect associated with the analogue  apparent horizon.
The surface gravity $\kappa$ of a Killing horizon can be defined by
\begin{equation}
\xi^\nu\nabla_\nu\xi_\mu=\kappa \xi_\mu ,
\label{eq223}
\end{equation}
evaluated on the horizon. 
If the geometry were stationary, the analogue apparent horizon would coincide with the 
analogue event horizon 
at the hypersurface defined by $v=c_\pi$.

In the case of non-stationary spacetime, the apparent horizon is neither Killing nor null.
The definition of surface gravity in this case is not unique
\cite{nielsen2}
 and
several ideas have been put forward how to generalize the definition of surface gravity for the apparent horizon
\cite{hayward2,hayward,fodor,mukohyama-booth}. We adopt the  prescription of 
\cite{hayward2} which, we believe, is most suitable for spherical symmetry. This prescription involves the so called Kodama vector $K^\mu$ \cite{kodama} which generalizes 
the concept of the time translation Killing vector to non-stationary spacetimes.
The Kodama vector we define as \cite{hayward,abreu}
\begin{equation}
K^\alpha= k \epsilon^{\alpha\beta}  n_\beta \;\; {\rm for}\; \; \alpha=0,1; \hspace{1cm}
K^i=0\;\; {\rm for} \;\; i=2,3,
\label{eq224}
\end{equation}
where $\epsilon^{\alpha\beta}$ is the covariant 
two-dimensional Levi-Civita tensor in the 
space normal to the surface of spherical symmetry
and  $n_\alpha$ is a vector normal to that  surface.
The normalization 
  factor $k$ has to be adjusted so that $K^\mu$
coincides with the time translation Killing
vector $\xi^\mu$  for a stationary geometry. 
In analogy with (\ref{eq223}) the surface gravity $\kappa$ is defined by  \cite{hayward2,hayward3}
\begin{equation}
K^\alpha \nabla_{[\alpha} K_{\beta]}=\kappa K_\beta ,
\label{eq227}
\end{equation}
where  the quantities should be evaluated on the trapping horizon.  
Using this definition we find \cite{bilic-tolic2}
\begin{equation}
\kappa =
\frac{c_\pi}{2\tau }
\frac{1 + 2c_\pi v (1-v)- (2+c_\pi) v^3}{\gamma v(1+c_\pi v )^2}
+\frac{\ddot{\alpha}}{2\beta}\frac{v}{\gamma (1+c_\pi v)^2} 
\label{eq231}
\end{equation}
where it is understood that the right-hand side is evaluated on the trapping horizon. 

In the limiting case 
when the quantities  $a$,  and  $c_\pi$ are constants,
 the apparent horizon
is determined  by the condition $v = c_\pi$
and the expression for $\kappa$  reduces to $\kappa =1/2t=\sqrt{1-c_\pi^2}/2\tau$. 
Hence, the analogue surface gravity is finite for any physical value of $c_\pi$ and is maximal when $c_\pi=0$.
However, with $c_\pi=0$ the horizon degenerates to a point located at the origin $r=0$. The temperature 
\begin{equation}
T_H=\frac{\kappa}{
 2\pi}
\label{eq044}
\end{equation}
 is
the analogue Hawking temperature of thermal pions emitted at the apparent horizon
as perceived  by an observer at
 infinity. Since the background geometry is flat, this temperature  equals
 the locally measured Hawking temperature at the horizon.
As we move along the trapping horizon the radius of the apparent horizon increases
and the Hawking  temperature decreases rapidly with $\tau$.
Hence, there is a correlation between $T_H$ and the local fluid temperature $T$,
which is related to $\tau$.
 
In contrast to the usual general relativistic Hawking effect, where the Hawking temperature 
is much smaller than the temperature of the background, the analogue horizon temperature 
is of the order or even larger than the local temperature of the fluid. 
The Hawking temperature correlates with the local temperature of the 
fluid at the apparent horizon and diverges at the critical point \cite{bilic-tolic1}.

\subsubsection*{Conclusion}

Formation of an analogue apparent horizon in an expanding hadronic fluid 
is similar to the formation of a black hole in a gravitational collapse
although the role of  an outer trapped surface is exchanged with that of 
an inner trapped surface.
Unlike a black hole in general relativity, the formation of which is indicated by the existence
of an outer marginally trapped surface,
the formation of an analogue black (or white) hole in an expanding fluid 
is indicated by the existence of a  future  or past inner marginally  trapped surface.

\subsection*{Acknowledgments}
This work was supported by the Ministry of Science,
Education and Sport
of the Republic of Croatia under Contract No. 098-0982930-2864.
and 
partially supported by the ICTP-SEENET-MTP grant PRJ-09 ``Strings and Cosmology`` 
in the frame of the SEENET-MTP Network. N.B. thanks CNPq, Brazil, for partial support and the University of Juiz de Fora where a part of
this work has been completed.


\end{document}